\documentstyle[11pt]{article}
\title{\bf The perspectives of decoding the nature of the
 \boldmath{$a_0(980)$} and \boldmath{$f_0(980)$} mesons and of defining the relative
phase between the three-gluon and one-photon amplitudes in
 the \boldmath{$J/\psi$} decays
 \thanks{ Invited talk given on March 1, 2003 at "KEK Workshop on hadron
 spectroscopy and chiral particle search in $J/\psi$ decay data at BES",
February 28 -  March 1, 2003, Tsukuba, Japan }  }
\author{N.N. Achasov\\[1cm]
 Laboratory of Theoretical Physics,\\
 Sobolev Institute for Mathematics,\\
  Academician Koptiug Prospekt, 4,\\
   Novosibirsk, 630090, Russia
 \thanks{ E-mail: achasov@math.nsc.ru}}
\date{\today}
\begin{document}
\maketitle
\begin{abstract}
  It is argued that the search of the
$J/\psi\to f_0(980)\omega$ and $J/\psi\to a_0(980)\rho$ decays and
the more precise definition of $B(J/\psi\to f_0(980)\phi)$ are the
urgent purposes in the $J/\psi$ spectroscopy.

 It is shown that the study of the
$\omega-\rho^0$ interference pattern in the $J/\psi\to
(\rho^0+\omega )\eta\to\pi^+\pi^-\eta$ decay provides  evidence
for the large (nearly $90^\circ$) relative phase between the
isovector one-photon  and  three-gluon decay amplitudes .
\end{abstract}

\section{Introduction}
As  is well known that the $J/\psi$ decays have played an
outstanding role in creation of Standard Model, including creation
QCD. But up to now  potentialities the $J/\psi$ decays  to give
top level results, that is new physics, are far from to be
exhausted as is clear from two topics considered below.

In Section 1 it is shown that there are good potentialities to
clear the nature of the scalar $f_0(980)$ and $a_0(980)$- mesons
studying the  $J/\psi\to f_0(980)\omega$, $J/\psi\to a_0(980)\rho$
and $J/\psi\to f_0(980)\phi$ decays.

In Section 2 it is shown that that there good potentialities to
get a relative phase between the isovector one-photon  and
three-gluon decay amplitudes in $J/\psi$ decays studying the
$J/\psi\to (\rho^0+\omega )\eta\to\pi^+\pi^-\eta$  and  $J/\psi\to
\omega\eta$ decays.

A brief summary is given in Section 3.

\section{The \boldmath{$J/\psi$} decays about the nature
of the scalar \boldmath{$f_0(980)$} and \boldmath{$a_0(980)$}
mesons}

\subsection{The \boldmath{$J/\psi$} decays about the nature
of the scalar \boldmath{$a_0(980)$} meson}

 The following data
is of very interest for our purposes:
\begin{eqnarray}
\label{a0rho} && B(J/\psi\to a_0(980)\rho) < 4.4\cdot 10^{-4}\ \ \
\ \mbox{\cite{kopke-89}\ \ and}
\end{eqnarray}
\begin{eqnarray}
\label{a2rho}
 && B(J/\psi\to a_2(1320)\rho)= (109\pm 22)\cdot 10^{-4}\ \ \ \
 \mbox{\cite{pdg-02}.}
\end{eqnarray}

The suppression
\begin{eqnarray}
\label{a0rho/a2rho}
 && B(J/\psi\to a_0(980)\rho) /B(J/\psi\to
a_2(1320)\rho)< 0.04\pm 0.008
\end{eqnarray}
seems strange, if one considers the $a_2(1320)$ and
$a_0(980)$-states as the tensor and scalar two-quark states from
the same P-wave multiplet with the quark structure
\footnote{ It
cannot be too highly stressed that in $J/\psi$ decays there is no
suppression of creation of isovector P-wave $q\bar q$ states in
comparison with creation of isovector S-wave $q\bar q$ states.
Please compare $B(J/\psi\to a_2(1320)\rho)= (109\pm 22)\cdot
10^{-4}$ with $B(J/\psi\to \pi\rho)= (127\pm 9)\cdot 10^{-4}$\ \
\cite{pdg-02}.}
\begin{eqnarray}
\label{a0qq}
 && a_0^0=(u\bar u-d\bar d)/\sqrt{2}\ \ ,\ \
a^+_0=u\bar d\ \ ,\ a^-_0 = d\bar u\,.
\end{eqnarray}

While the four-quark nature of the $a_0(980)$-meson with the
symbolic quark structure, similar (but not identical) the MIT-bag
state \cite{jaffe-77},
\begin{eqnarray}
\label{a0qqqq}
 && a_0^+(980)= us\bar d\bar s\,,\ a_0^0(980)=
\frac{(us\bar u\bar s-ds\bar d\bar s)}{\sqrt{2}}\,,\ a_0^-(980)=
ds\bar u\bar s\,.
\end{eqnarray}
is not c®ntrary to the suppression in Eq. (\ref{a0rho/a2rho}).

 {\bf So, the improvement of the upper limit (\ref{a0rho}) and the
search for the \boldmath{$J/\psi\to a_0(980)\rho$} decays  are the
urgent purposes in the study of the \boldmath{$J/\psi$} decays!}

\subsection{The  \boldmath{$J/\psi$} decays about the nature
of the scalar \boldmath{$f_0(980)$} meson}

Let us discuss  a possibility to treat the $f_0(980)$-meson as the
quark-antiquark state.

The hypothesis that the $f_0(980)$-meson is the lowest two-quark
P-wave scalar state with the quark structure
\begin{eqnarray}
\label{f0qq}
 && f_0=(u\bar u+d\bar d)/\sqrt{2}
\end{eqnarray}
contradicts the following facts:

i) the weak coupling with gluons \cite{eigen-88}
\begin{eqnarray}
\label{f0gluons} && B(J/\psi\to\gamma f_0(980)\to\gamma\pi\pi) <
1.4\cdot 10^{-5}
\end{eqnarray}
opposite the expected one \cite{farrar-94} for Eq. (\ref{f0qq})
\begin{eqnarray}
\label{farrar2q}
 && B(J/\psi\to\gamma f_0(980))\geq
B(J/\psi\to\gamma f_2(1270))/4\simeq (3.45\pm 0.35)\cdot
10^{-4}\,;
\end{eqnarray}

ii) the decays $J/\psi\to f_0(980)\omega$, $J/\psi\to
f_0(980)\phi$, $J/\psi\to f_2(1270)\omega$ and $J/\psi\to
f_2'(1525)\phi$\ \ \cite{pdg-02}:
\begin{eqnarray}
\label{f0omega} && B(J/\psi\to f_0(980)\omega)=(1.4\pm 0.5)\cdot
10^{-4}\,,
\end{eqnarray}
\begin{eqnarray}
\label{f0phi} && B(J/\psi\to f_0(980)\phi)=(3.2\pm 0.9)\cdot
10^{-4}\,,
\end{eqnarray}
\begin{eqnarray}
\label{f2omega} && B(J/\psi\to f_2(1270)\omega)=(4.3\pm 0.6)\cdot
10^{-3}\ \ \mbox{and}
\end{eqnarray}
\begin{eqnarray}
\label{f2'phi} && B(J/\psi\to f_2'(1525)\phi)=(8\pm 4)\cdot
10^{-4}\,.
\end{eqnarray}

The suppression
\begin{eqnarray}
\label{f0omega/f2omega} && B(J/\psi\to f_0(980)\omega)
/B(J/\psi\to f_2(1270)\omega)= 0.033\pm 0.013
\end{eqnarray}
looks strange in the model under consideration as well as Eq.
(\ref{a0rho/a2rho}) in the model (\ref{a0qq})
 \footnote{ It cannot
be too highly stressed that in $J/\psi$ decays there is also no
suppression of creation of isoscalar P-wave $q\bar q$ states in
comparison with creation of isoscalar S-wave $q\bar q$ states.
Please compare $B(J/\psi\to f_2(1270)\omega)=(4.3\pm 0.6)\cdot
10^{-3}$ with $B(J/\psi\to \eta\omega)= (1.58\pm 0.16)\cdot
10^{-3}$\ \ \cite{pdg-02}.}.

I would like to emphasize that from my point of view the DM2
Collaboration did not observed the $J/\psi\to f_0(980)\omega$
decay and should give a upper limit instead of Eq.
(\ref{f0omega}).

{\bf So, the search for the \boldmath{$J/\psi\to f_0(980)\omega$}
decay is the urgent purpose in the study of the
\boldmath{$J/\psi$} decays!}

The existence of the $J/\psi\to f_0(980)\phi$ decay of greater
intensity than the $J/\psi\to f_0(980)\omega$ decay ( compare Eq.
(\ref{f0omega}) and Eq. (\ref{f0phi}) ) shuts down the model
(\ref{f0qq}) for in the case under discussion the $J/\psi\to
f_0(980)\phi$-decay should be strongly suppressed in comparison
with the $J/\psi\to f_0(980)\omega$-decay by the OZI-rule.

{\bf So, Eq. (\ref{f0qq}) is excluded at a level of physical
rigor.}

Can one consider the $f_0(980)$-meson as the near $s\bar s$-state?

It is impossible without a gluon component. Really, it is
anticipated for the scalar $s\bar s$-state from the lowest P-wave
multiplet that \cite{farrar-94}
\begin{eqnarray}
\label{farrar2s} && B(J/\psi\to\gamma f_0(980))\geq
B(J/\psi\to\gamma f_2^\prime(1525))/4 \simeq 1.6\cdot 10^{-4}
\end{eqnarray}
opposite Eq. (\ref{f0gluons}), which requires properly that the
$f_0(980)$-meson to be the 8-th component of the $SU_f(3)$-oktet
\begin{eqnarray}
\label{oktet} && f_0(980)=(u\bar u+d\bar d-2s\bar s)/\sqrt{6}\,.
\end{eqnarray}

This structure gives

\begin{eqnarray}
\label{f0phif0omega} &&B(J/\psi\to f_0(980)\phi)=(2\lambda\approx
1)\cdot B(J/\psi\to f_0(980) \omega)\,,
\end{eqnarray}
which is on the verge of conflict with experiment, compare Eq.
(\ref{f0omega}) with Eq. (\ref{f0phi}). Here $\lambda$ takes into
account the strange sea suppression.

Equation (\ref{oktet}) contradicts also the strong coupling with
the $K\bar K$-channel \cite{achasov-84,achasov-8997}
\begin{eqnarray}
\label{r} && 1<R=|g_{f_0K^+K^-}/g_{f_0\pi^+\pi^-}|^2\leq 10
\end{eqnarray}
 for the prediction
\begin{eqnarray}
\label{roktet}
 &&R =|g_{f_0K^+K^-}/g_{f_0\pi^+\pi^-}|^2=(\sqrt{\lambda}-2)^2/4\simeq
0.4\,.
\end{eqnarray}

In addition,  the mass degeneration $m_{f_0}\simeq m_{a_0}$ is
coincidental in this case if to treat the $a_0$-meson as the
four-quark state or contradicts the light hypothesis (\ref{a0qq}).

The introduction of a gluon component, $gg$, in the
$f_0(980)$-meson structure  allows the puzzle of weak coupling
with two gluons (\ref{f0gluons}) and with two photons but the
strong coupling with the $K\bar K$-channel  to be resolved easy
\cite{achasov-98}:
\begin{eqnarray}
\label{f0ss} && f_0=gg\sin\alpha +\left [\left (1/\sqrt{2}\right
)(u\bar u+d\bar d)
\sin\beta + s\bar s\cos\beta\right ]\cos\alpha\,,\nonumber\\
&&\tan\alpha=-O(\alpha_s)\left (\sqrt{2}\sin\beta +\cos\beta\right
)\,,
\end{eqnarray}
where $\sin^2\alpha\leq 0.08$ and $\cos^2\beta > 0.8$.

So, the $f_0(980)$-meson is near the $s\bar s$-state, as
in\cite{nils-82} .

It gives
\begin{eqnarray}
\label{f0omega/f0phiss} &&0.1 < \frac{B(J/\psi\to
f_0(980)\omega)}{B(J/\psi\to f_0(980)\phi)}=
\frac{1}{\lambda}\tan^2\beta < 0.54\,.
\end{eqnarray}
As for the experimental value,
\begin{eqnarray}
\label{f0omega/f0phiexp}
 && B(J/\psi\to f_0(980)\omega)/B(J/\psi\to f_0(980)\phi)=0.44\pm 0.2\,,
\end{eqnarray}
it needs refinement.

{\bf Remind that in my opinion  the \boldmath{$J/\psi\to
f_0(980)\omega$} was not observed!}

The scenario with the $f_0(980)$ meson as in Eq. (\ref{f0ss}) and
with the $a_0(980)$ meson as
the two-quark state (\ref{a0qq}) runs into following difficulties:\\
i) it is impossible to explain the $f_0$ and $a_0$-meson mass degeneration in a natural way;\\
ii) it is possible to get only \cite{achasov-8997}
\begin{eqnarray}
\label{phigammaf0a0} && B(\phi\to\gamma
f_0\to\gamma\pi^0\pi^0)\simeq 1.7\cdot 10^{-5}\,,
\nonumber\\
&& B(\phi\to\gamma a_0\to\gamma\pi^0\eta)\simeq 10^{-5}\,.
\end{eqnarray}
iii) it is also predicted
\begin{eqnarray}
\label{a0rhof0phi} && B(J/\psi\to a_0(980)\rho)=(3/\lambda\approx
6)\cdot B(J/\psi\to f_0(980)\phi)\,,
\end{eqnarray}
that has almost no chance, compare Eqs. (\ref{a0rho}) and
(\ref{f0phi}).

Note that the $\lambda$ independent prediction
\begin{eqnarray}
\label{f0phi/f2'phia0rho/a2rho}
&& B(J/\psi\to f_0(980)\phi)/B(J/\psi\to f_2'(1525)\phi)=\nonumber\\
&& = B(J/\psi\to a_0(980)\rho)/B(J/\psi\to a_2(1320)\rho)
\end{eqnarray}
is excluded by the central figure in
\begin{eqnarray}
\label{f0phi/f2'phi} && B(J/\psi\to f_0(980)\phi) /B(J/\psi\to
f_2'(1525)\phi)= 0.4\pm 0.23\,,
\end{eqnarray}
obtained from Eqs. (\ref{f0phi}) and (\ref{f2'phi}), compare with
Eq. (\ref{a0rho/a2rho}). But, certainly, experimental error is too
large.

{\bf Even twofold increase in accuracy of measurement of Eq.
(\ref{f0phi/f2'phi}) could be crucial in the fate of the scenario
under discussion.}

The prospects for the model of the $f_0(980)$-meson as the almost
pure $s\bar s$-state (\ref{f0ss}) and the $a_0(980)$-meson as the
four-quark state (\ref{a0qqqq}) with the coincidental mass
degeneration is rather poor especially as  the mechanism without
creation  and annihilation of the additional $u\bar u$ pair, i.e.
the OZI-superallowed $\left (N_C\right )^0$ order transition $\phi
= s\bar s\to\gamma s\bar s = \gamma f_0(980)$
 \footnote{ In this regard the $\left
(N_C\right )^0$ order mechanism is similar to the principal
mechanism of the $\phi\to\gamma\eta '(958)$ decay ($\phi = s\bar
s\to\gamma s\bar s = \gamma\eta'(958)$).},
 cannot explain the
photon spectrum in $\phi\to\gamma f_0(980)\to\gamma\pi^0\pi^0$,
which requires the domination of the $K^+K^-$ intermediate state
in the $\phi\to\gamma f_0(980)$ amplitude: $\phi\to
K^+K^-\to\gamma f_0(980)$, as is shown in Refs.
\cite{achasov-01,achasov-010203}! The $\left (N_C\right )^0$ order
transition
 is bound to have a small weight in the
large $N_C$ expansion of the $\phi = s\bar s\to\gamma f_0(980)$
amplitude, because this term does not contain the $K^+K^-$
intermediate state, which emerges only in the next to leading term
of the $1/N_C$ order, i.e., in the OZI forbidden transition
\cite{achasov-010203}.

While the four-quark model with the symbolic structure
\begin{eqnarray}
\label{f0qqqq}
 && f_0(980) = \frac{(us\bar u\bar s+ds\bar d\bar s)}{\sqrt{2}}\cos\theta + ud\bar u\bar d\sin\theta\,,
\end{eqnarray}
similar (but not indentical1) the MIT-bag state \cite{jaffe-77},
reasonably justifies all unusual features of the $f_0(980)$-meson
\cite{achasov-84,achasov-9188,achasov-98,achasov-010203}.

\section{ The \boldmath{$\omega-\rho^0$} interference pattern in the
\boldmath{$J/\psi\to (\rho^0+\omega )\eta\to\pi^+\pi^-\eta$} decay
about the relative phase between the three-gluon and one-photon
amplitudes in the \boldmath{$J/\psi$} decays}

 In the last few years it  has been noted that the
single-photon and three-gluon amplitudes in the two-body
$J/\psi\to 1^-0^-$ and $J/\psi\to 0^-0^-$
\cite{castro-95,suzuki-98,suzuki-99}  decays appear to have
relative phases nearly $90^\circ$.

This unexpected result is very important to the observability of
CP violating decays  as well as to the nature of the $J/\psi\to
1^-0^-$ and $J/\psi\to 0^-0^-$ decays
\cite{castro-95,suzuki-98,suzuki-99,rosner-99,tuan-99,gerard-99,tuan-01}.
In particular, it points to a non-adequacy of their description
built upon the perturbative QCD, the hypothesis of the
factorization of short and long distances, and  specified wave
functions of final hadrons.  Some peculiarities of electromagnetic
form factors in the $J/\psi$ mass region were discussed in Ref.
\cite{achasov-1998}.

The analysis \cite{castro-95,suzuki-98,suzuki-99} involved
theoretical assumptions relying on the strong interaction
$SU_f(3)$-symmetry, the strong interaction $SU_f(3)$-symmetry
breaking and the $SU_f(3)$ transformation properties of the
one-photon annihilation amplitudes. Besides,
 effects of the $\rho-\omega$ mixing in the
$J/\psi\to 1^-0^-$ decays were not taken into account in Ref.
\cite{castro-95} while
 in Ref. \cite{suzuki-98} the $\rho-\omega$ mixing was taken into
 account incorrectly , see the discussion in Ref. \cite{pisma}.
  Because of this, the model independent
determination of these phases are required.

Fortunately, it is possible to check the conclusion of Refs.
\cite{castro-95,suzuki-98} at least in one case
\cite{pisma,physrev}. We mean the relative phase between the
amplitudes of the one-photon $J/\psi\to\rho^0\eta$ and three-gluon
$J/\psi\to\omega\eta$ decays.

The point is that the $\rho^0-\omega$ mixing amplitude is
reasonably well studied
\cite{goldhaber-69,gourdin-69,renard-70,achasov-78,achasov-92,achasov-74,pdg-98}.
Its module and phase are known. The module of the ratio of the
amplitudes of the $\rho$ and $\omega$ production can be obtained
from the data on the branching ratios of the $J/\psi$-decays. So,
the investigation of the $\omega-\rho$ interference in the
$J/\psi\to (\rho^0+\omega )\eta\to\rho^0\eta\to\pi^+\pi^-\eta$
decay provides a way of measuring the relative phase of the
$\rho^0$ and $\omega$ production amplitudes.

Indeed, the $\omega-\rho$ interference pattern in the $J/\psi\to
(\rho^0+\omega )\eta\to\rho^0\eta\to\pi^+\pi^-\eta$ decay is
conditioned by the $\rho^0-\omega$ mixing and the ratio of the
amplitudes of the $\rho^0$ and $\omega$ production:
\begin{eqnarray}
\label{spectrum1} && \frac{dN}{dm}= N_\rho (m)\frac{2}{\pi}m\Gamma
(\rho\to\pi\pi\,,\, m)\times\nonumber\\
&&\left |\frac{1}{D_\rho(m)}\left (1-\varepsilon (m)\left
[\frac{N_\omega (m)} {N_\rho (m)}\right
]^{\frac{1}{2}}\exp\left\{i\left (\delta_\omega -
 \delta_\rho\right )\right\}\right )+ \right. \nonumber\\[1pc]
&& \left. +\ \ \frac{1}{D_\omega (m)}\left (\varepsilon (m)+
g_{\omega\pi\pi}/g_{\rho\pi\pi}\right )\left [\frac{N_\omega
(m)}{N_\rho (m)}\right ]^{\frac{1}{2}} \exp\left\{i\left
(\delta_\omega - \delta_\rho\right )\right\}\right |^2
\end{eqnarray}
with
\begin{eqnarray}
\label{epsilon} && \varepsilon
(m)=-\frac{\Pi_{\omega\rho^0}(m)}{m_\omega^2-m_\rho^2 + im\left
(\Gamma_\rho (m)-\Gamma_\omega (m)\right )},
\end{eqnarray}
where m is the invariant mass of the $\pi^+\pi^-$-state, $N_\rho
(m)$ and $N_\omega (m)$ are the squares of the modules of the
$\rho$ and $\omega$ production amplitudes, $\delta_\rho$ and $
\delta_\omega$ are their phases, $\Pi_{\omega\rho^0}(m)$ is the
amplitude of the $\rho-\omega$ transition, $D_V(m)=m_V^2 - m^2 -
im\Gamma_V(m)$, $V=\rho$, $\omega$.

We obtained in Refs. \cite{pisma,physrev}
 \footnote{ If we use Ref.
\cite{pdg-02} we shall obtain $ \varepsilon (m_\omega)+
g_{\omega\pi\pi}/g_{\rho\pi\pi}= ( 2.99\pm 0.25 )\cdot
10^{-2}\exp\left \{i\left ( 102\pm 1\right )^\circ\right \}$.}
\begin{eqnarray}
\label{epsiloneff} && \varepsilon (m_\omega)+
g_{\omega\pi\pi}/g_{\rho\pi\pi}= ( 3.41\pm 0.24 )\cdot
10^{-2}\exp\left \{i\left ( 102\pm 1\right )^\circ\right \}\,.
\end{eqnarray}

The branching ratio of the $\omega\to\pi\pi$ decay
\begin{eqnarray}
\label{b} && B\left (\omega\to\pi\pi\right )=\frac{\Gamma \left
(\rho\to\pi\pi\,,\, m_\omega\right )}{\Gamma_\omega
(m_\omega)}\cdot\left |\varepsilon (m_\omega) +
g_{\omega\pi\pi}/g_{\rho\pi\pi}\right |^2.
\end{eqnarray}

The data \cite{markiii-88,dm2-90} were fitted with the function
\begin{eqnarray}
\label{fit} N(m) = L(m) + \left |\left (N_\rho\right
)^{\frac{1}{2}}F_\rho^{BW}(m) + \left (N_\omega\right
)^{\frac{1}{2}}F_\omega^{BW}(m)\exp\{i\phi\} \right |^2\,,
\end{eqnarray}
where $F_\rho^{BW}(m)$ and $F_\omega^{BW}(m)$ are the appropriate
Breit-Wigner terms \cite{markiii-88} and $L(m)$ is a polynomial
background term.

The results are
\begin{eqnarray}
\label{experiment}
 && \phi = ( 46 \pm 15)^\circ\,,\quad
N_\omega(m_\omega)/N_\rho =
8.86\pm 1.83\ \ \mbox{\cite{markiii-88}}\,,\nonumber\\
&& \phi = - 0.08\pm 0.17=(-4.58\pm 9.74)^\circ\,,\quad
N_\omega(m_\omega)/N_\rho =7.37\pm 1.72\ \ \mbox{\cite{dm2-90}}\,.
\end{eqnarray}

From Eqs. (\ref{spectrum1}), (\ref{b}), and (\ref{fit}) it follows
\begin{eqnarray}
\label{Nrho}
 && N_\rho = N_\rho (m_\rho)\left |1 - \varepsilon
(m_\rho)\left [ N_\omega(m_\rho)/N_\rho(m_\rho)\right
]^{\frac{1}{2}} \exp\{i\left(\delta_\omega - \delta_\rho\right
)\}\right |^2\,,\\
 \label{Nomega}
  && N_\omega =
B(\omega\to\pi\pi)N_\omega(m_\omega)\,,\\
\label{phi} && \phi = \delta_\omega - \delta_\rho + arg\left
[\varepsilon (m_\omega)+ g_{\omega\pi\pi}/g_{\rho\pi\pi} \right ]
- \nonumber \\ && - arg\left\{1 - \varepsilon (m_\rho)\left
[N_\omega(m_\rho)/N_\rho(m_\rho)\right ]^{\frac{1}{2}}
\exp\{i\left(\delta_\omega - \delta_\rho\right )\} \right\}\simeq
\nonumber\\ && \simeq \delta_\omega - \delta_\rho + arg\left
[\varepsilon (m_\omega)+ g_{\omega\pi\pi}/g_{\rho\pi\pi}\right ]
-\nonumber \\ &&- arg\left\{1 - \left |\varepsilon
(m_\omega)\right |\left [N_\omega(m_\omega)/N_\rho \right
]^{\frac{1}{2}}\exp\{i\phi\} \right\}\,.
\end{eqnarray}

From Eqs. (\ref{epsiloneff}), (\ref{experiment}) and (\ref{phi})
we get that
\begin{eqnarray}
\label{markiii} && \delta_\rho - \delta_\omega =% \delta_\gamma =
(60\pm 15 )^\circ\ \
\mbox{\cite{markiii-88}\ \ and}\\
\label{dm2} && \delta_\rho - \delta_\omega = %\delta_\gamma =
(106\pm 10 )^\circ\ \ \mbox{\cite{dm2-90}}\,.
\end{eqnarray}

Whereas $\delta_\rho$ is the phase of the isovector one-photon
amplitude, $\delta_\omega$ is the phase of the sum of the
three-gluon amplitude and the isoscalar one-photon amplitude. But
luckily for us the latter is a small correction. Really, it
follows from the structure of the electromagnetic current
\begin{eqnarray}
\label{current}
 && j_\mu (x) =\frac{2}{3}\bar u(x)\gamma_\mu u(x) -
\frac{1}{3}\bar d(x)\gamma_\mu d(x) - \frac{1}{3}\bar
s(x)\gamma_\mu s(x) + ...
\end{eqnarray}
and the Okubo-Zweig-Iizuka rule the ratio for the amplitudes under
consideration ({\bf please image all possible diagrams!}):
\begin{eqnarray}
\label{sv}
 && \frac{A\left (J/\psi\to\mbox{ the isoscalar
photon}\to\omega\eta\right )}{A\left (J/\psi\to\mbox{ the
isovector photon}\to\rho\eta\right )\equiv A\left
(J/\psi\to\rho\eta\right )}=\frac{1}{3}\,.
\end{eqnarray}
Taking into account Eqs. (\ref{experiment}) and (\ref{Nrho}) one
gets
\begin{eqnarray}
\label{sg}
 && \frac{\left|A\left (J/\psi\to\mbox{ the isoscalar
photon}\to\omega\eta\right )\right|}{\left|A\left (J/\psi\to\mbox{
the three-gluon}\to\omega\eta\right )\right
|}\approx\frac{1}{9}\,.
\end{eqnarray}
From Eqs. (\ref{markiii}), (\ref{dm2}) and (\ref{sg}) one gets
easily for the relative phase ($\delta$) between the isovector
one-photon and  three gluon decay amplitudes
\begin{eqnarray}
\label{markiii1} && \delta = (60\pm 15 )^\circ\ - 4^\circ\ \ \ \ \
\mbox{\cite{markiii-88}\ \  and}\\
\label{dm21} && \delta = (106\pm 10 )^\circ - 6^\circ\ \ \ \ \
\mbox{\cite{dm2-90},}
\end{eqnarray}
if the isovector and isoscalar  one-photon decay amplitudes have
the same phase. In case the isoscalar one-photon and  three-gluon
(isoscalar also!) decay amplitudes have the same phase
\begin{eqnarray}
\label{markiii2} && \delta = (60\pm 15 )^\circ\ \ \ \ \ \ \ \
\mbox{\cite{markiii-88}\ \ and}\\
\label{dm22} && \delta = (106\pm 10 )^\circ\ \ \ \ \ \ \
\mbox{\cite{dm2-90}.}
\end{eqnarray}

So, both the MARK III Collaboration \cite{markiii-88} and the DM2
Collaboration \cite{dm2-90}, see Eqs. (\ref{markiii1}),
(\ref{markiii2}) and (\ref{dm21}), (\ref{dm22}), provide support
for the large (nearly $90^\circ$) relative phase between the
isovector one-photon  and three-gluon decay amplitudes.

The DM2 Collaboration used statistics only half as high as the
MARK III Collaboration, but, in contrast to the MARK III
Collaboration, which fitted $N_\omega$ as a free parameter, the
DM2 Collaboration calculated it from the branching ratio of
$J/\psi\to\omega\eta$ using Eq. (\ref{Nomega}).

In summary I should emphasize that it is urgent to study this
fundamental problem once again with KEDR in Novosibirsk and with
BES II in Beijing.

But I am afraid that only the $\tau$-CHARM factory could solve
this problem in the exhaustive way.

\section{Conclusion}

 {\bf So, the search for the \boldmath{$J/\psi\to a_0(980)\rho$} and
\boldmath{$J/\psi\to f_0(980)\omega$} decays, the more precise
definition of \boldmath{$B(J/\psi\to f_0(980)\phi)$}, and the
study of the \boldmath{$\omega-\rho^0$} interference pattern in
the \boldmath{$J/\psi\to (\rho^0+\omega )\eta\to\pi^+\pi^-\eta$}
decay are the urgent purposes in the \boldmath{$J/\psi$}
spectroscopy!}

\section*{Acknowledgements}

 I would like to thank Professor Ishida-San, Professor Takamatsu-San and Professor Tsuru-San very much for the
 Invitation, the financial support and the generous hospitality.

 This work was supported in part by RFBR, Grant No 02-02-16061.

\end{document}